\documentclass[usenatbib,onecolumn]{mnras}

\usepackage{graphicx}
\usepackage{wrapfig}
\usepackage{natbib}

\newcommand{\be}{\begin{equation}} 
\newcommand{\ee}{\end{equation}}
\newcommand{\nn}{\mbox{} \nonumber \\ \mbox{} }
\newcommand{\ba}{\begin{eqnarray}}
\newcommand{\ea}{\end{eqnarray}}
\newcommand{\om}{\omega}

\newcommand\eg{\textit{e.g.,\ }}

\newcommand{\Bf}{{magnetic field}}
\newcommand{\Bfs}{{magnetic fields}}

\newcommand{\NS}{neutron star}

\newcommand{\EM}{electromagnetic}

\newcommand{\Lf}{Lorentz factor}

\title{On the nature of Fast Blue Optical Transients}

\author{Maxim Lyutikov\\
Department of Physics and Astronomy, Purdue University, 
 525 Northwestern Avenue,
West Lafayette, IN
47907-2036}

\begin{document}

\maketitle

\begin{abstract}
Short rise times of Fast Blue Optical Transients (FBOTs) require very light ejected envelopes, $M_{ej} \leq 10^{-1} M_\odot$, much smaller than of a  typical supernova. Short peak times also mean that FBOTs should be  hydrodynamically, not radioactively  powered. The detection by Chandra of X-ray  emission in AT2020mrf of $L_X \sim 10^{42} $ erg s$^{-1}$ after 328 days  implies total, overall dominant,  X-ray energetics   at the Gamma Ray Bursts (GRBs) level  of  $\sim   6 \times 10^{49}$ erg.    FBOTs  show  no evidence of relativistic motion, hence no beaming: the observed X-ray  luminosity  is similar to the true isotropic luminosity. 
  
 We further develop a model of Lyutikov \& Toonen (2019), whereby FBOTs are the results of a late  accretion  induced collapse (AIC)  of the product of  super-Chandrasekhar double white dwarf  (WD) merger between ONeMg  WD and another WD.
 Small ejecta mass, and the rarity of FBOTs, result from the  competition between mass loss from the merger product to the wind, and ashes  added to the core, on time scale of $\sim 10^3-10^4$ years. FBOTs occur only when the  envelope mass before AIC is $\leq  10^{-1} M_\odot$.   FBOTs proper come from central engine-powered radiation-dominated  forward shock as it propagates  through  ejecta.  FBOTs' duration is determined by the   diffusion time of photons   produced  by the NS-driven forward shock  within the expanding ejecta. 
 All the photons produced by the central source deep inside the ejecta escape almost simultaneously, producing a short bright event, violating the  ``Arnett's law''.   The high energy emission is generated at the highly relativistic and highly magnetized    termination shock, qualitatively similar to Pulsar Wind Nebulae. The X-ray bump observed in  AT2020mrf by SRG/eROSITA,   predicted by Lyutikov \& Toonen (2019),  is coming from the break-out of the engine-powered shock from the ejecta into the preceding wind. The model requires total energetics of  just few $\times 10^{50}$ ergs, slightly  above the observed X-rays. We predict that the system is hydrogen poor. 
  \end{abstract}

Keywords{(stars:) white dwarfs; (stars:) supernovae: general; stars: neutron}

\section{Introduction} 
Fast-rising blue optical transients \citep[FBOTs,][]{2014ApJ...794...23D} is a class of bright, short supernova explosions. AT2018cow \citep{2018ApJ...865L...3P,2019ApJ...871...73H,2019MNRAS.484.1031P,2019ApJ...872...18M}  is an exemplary event.  AT2018lqh \citep{2021ApJ...922..247O} and AT2020xnd  \cite{2021arXiv211005490H} had similar properties.   Recent observations of AT2020mrf  
\citep{2021arXiv211200751Y} further constraint the properties of the FBOTs' progenitors.  Similarity between AT2018cow, AT2018lqh, AT2020xnd and  AT2020mrf imply  similar physical system, with some variations of the parameters \citep[see][for a classification of rapid optical transients - most are classified as core-collapse events]{2021arXiv210508811H}.

The new constraint provided by the AT2020mrf is the relatively bright X-ray emission detected nearly a year after the original explosion.
Let us give estimates of the different  observed channels in FBOTs, taking  AT2020mrf as an example \citep{2021arXiv211200751Y} ($E_{v,r,mm,X} $ below is the total energetics in optical, radio, millimeter and X-rays) 
\begin{itemize}
\item  Optical: $M_v = -20$ for  3.7days  at   $3 \times 10^{43} $ erg/sec, $E_v= 10^{49} $ erg.
 \item Radio emission:  $\nu F_\nu = 1.2 \times10^{39}  erg /s$  at 261 days:  $E_r = 2.7 \times 10^{46} $ erg. 
\item Millimeter   \citep[Table 3 of ][]{2021arXiv211200751Y}:    $50 \mu $Jy at  16 GHz at 417.5 days:  $E_{mm}  \sim 1.5 \times 10^{47}$ erg. 
\item   X-rays: (i)  $2 \times 10 ^{43} erg s^{-1} $ at  20 days, $E_{\rm X} ^{(\rm early)}  =3.5 \times  10^{49}$ erg; (ii)   $10 ^{42} erg s^{-1} $ at  328 days, $E_{\rm X} ^{(\rm late)}  = 2.8 \times  10^{49}$ erg. \citep[The early  X-ray luminosity  of  AT2018cow  ($10 ^{43} erg s^{-1} $  at 20days,   $E_{\rm X} ^{(\rm early)}  = 1.7 \times  10^{49}$ erg.][]{2018ApJ...865L...3P}.
\end {itemize}

Thus,  the most energetically  constraining observation  of FBOTs is the high energy X-ray emission. Especially surprising  is the detection by {\it Chandra} of GRB-like emission of $L_X \sim 10^{42} $ erg s$^{-1}$ after 328 days  \citep[AT2020xnd had a similar flux at $\sim 50$ days][]{2021MNRAS.508.5138P}.    Optical  emission of FBOT proper comes close. 
\citep[The nature of the hot and luminous source detected by ][in the direction of  AT2018cow is  still uncertain.]{2022MNRAS.512L..66S}

The  implied total X-ray  energetics for AT2020mrf  is $\sim  6  \times 10^{49}$ erg s$^{-1}$.  Thus, 
    energies of AT2020mrf matches those of GRBs; but unlike GRBs they do not show relativistic velocities, hence their   luminosities/energetics    are of the order of the  true luminosities, while in GRBs true luminosities are smaller by the beaming factor, $\sim 10^{-2} $.  

Explaining the power and the photon energies  a year after an explosion   is the most challenging. In what follows we demonstrate that the model of \cite{2019MNRAS.487.5618L} both predicted the X-ray bump observed by  SRG /eROSITA,  can account for the total energetics, and   generally explains all the observed phenomena of FBOTs.

We also mention models of FBOTs by
\cite{2020ApJ...903...66L,2022MNRAS.513.3810G,2022RAA....22e5010S}. All these models rely on massive  hydrogen-rich stars. To keep the energy budgets under control the models require  highly jetted, GRB-like outflows. Explaining late X-ray emission, after nearly a year, is most challenging within these models. The main observations distinction is that the present model  in contrast advocates light hydrogen-poor ejecta.

\section{Hydrodynamic and radioactive  contributions to SN light-curves}

Supernova light curves is a complicated combination of hydrodynamic/internal heat dissipation  \citep{1971Ap&SS..10...28G} and radioactive  decay \citep{1969ApJ...157..623C}, see also  \cite{1997ARA&A..35..309F,2002RvMP...74.1015W}. The SN-Ia light curves are powered  mostly by  $^{56}$Ni  beta-decay of $\sim 0.5 M_\odot$ \citep{1982ApJ...253..785A,2000ApJ...530..744P}.   SN-Ib/c and SN-II are powered by a combination of shock heating, recombination of hydrogen \citep{1976Ap&SS..44..429G}, and later  by the
 $\rm{Co}{56}\rightarrow\rm{Fe}{56}$ decay \citep[see reviews by][]{2005IJMPA..20.6597N,2009ARA&A..47...63S,2018PhR...736....1L}.  ``FBOTs proper'' \footnote{We call ``FBOTs proper''  the short, few days, bright optical transients}, with the  short rise time, cannot be powered by  the radioactive decay \citep[][yet the  long term properties,  on the scale of $\sim$ months, can be/are affected]{2019MNRAS.487.5618L, 2022NatAs...6..249P}.  FBOTs proper must be hydrodynamically-powered. 
 
 Let us take an extreme position, and neglect the energy contribution from the  radioactive decay. It is overall  mildly significant, but comes at a later time, \S \ref{Contributionfromradioactivity}.
 
 Hydrodynamically powered light curves, with short rise time,  require small mass of the ejecta, \S \ref{small}, otherwise most of  the internal heat  or internal  shock power is lost to the adiabatic expansion. The required ejecta mass is $M_{ej} \leq 0.1 M_\odot$. This is an order of magnitude smaller than a typical SN-Ia/b/c or SN-II  ejecta. Alternative possibility - massive and very fast ejecta moving nearly with the speed of light  - requires enormous energy budget, on par with GRBs. 

\begin{figure} 
\includegraphics[width=.99\linewidth]{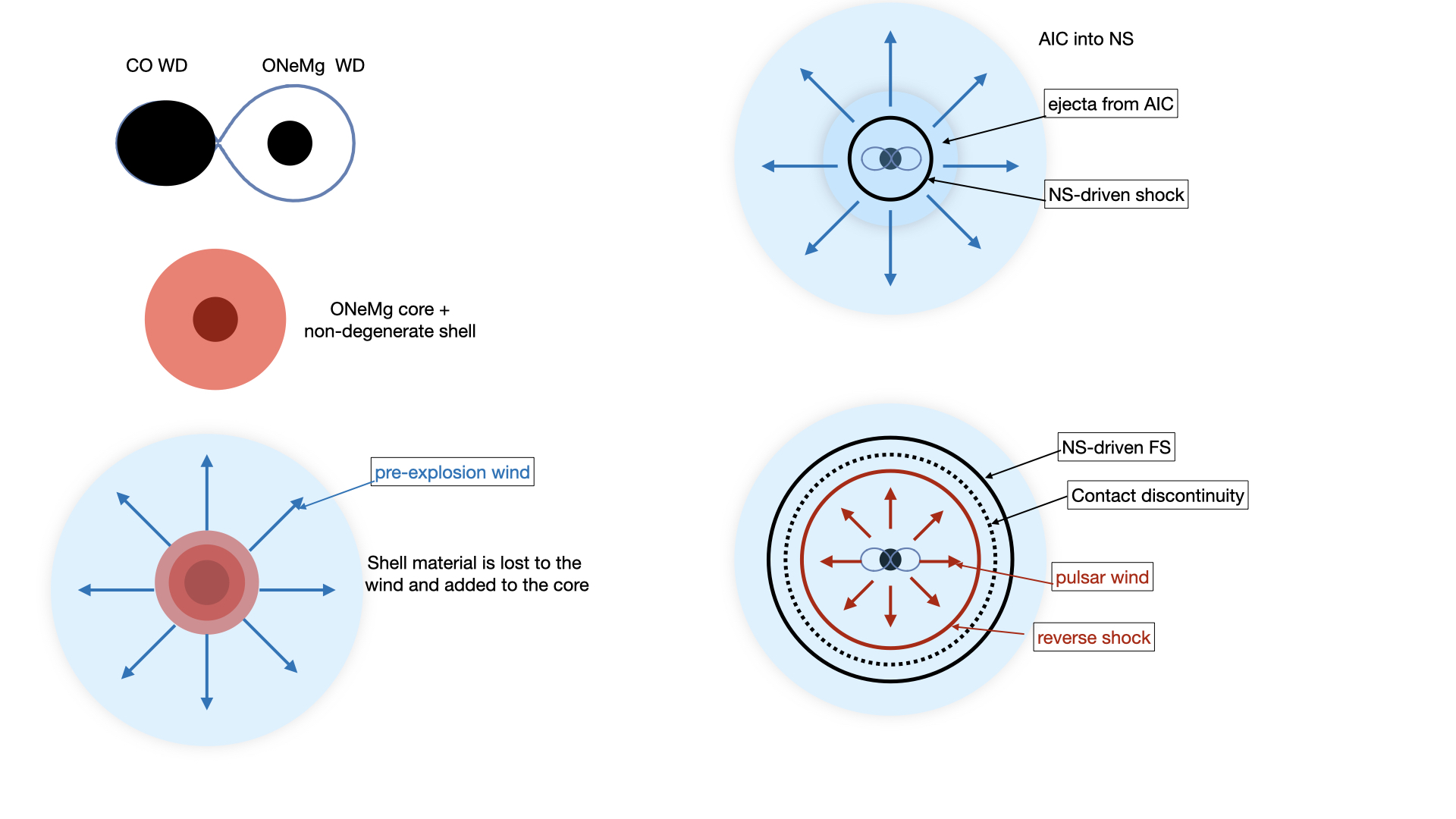}
\caption{Outline of the model of  \protect\cite{2019MNRAS.487.5618L}. A binary system,  via two separate evolutionary channels,  leads to the merger of a heavy ONeMg WD with another WD. The merger produce consists of a   non-degenerate envelope around the  ONeMg core.  Shell material is both lost to strong winds with  $\dot{M} \sim 10^{-5} -  10^{-3}  M_\odot$/yr, and  ashes added to the core \protect\cite[IRAS 00500+6713][ is currently at this stage, see \S \ref{Transits}]{2020A&A...644L...8O}.  The core experiences AIC \citep{1991ApJ...367L..19N}, producing fast rotating \NS\ and   ejecting   a  light remaining  shell of $M_{ej} \sim 10^{-2} - {\rm few} \, 0.1 M_\odot$. The secondary wind from the \NS\  propagates first through the ejecta  (the radiation-dominated forward shock produces FBOTs at this stage) and  breaks out into the wind  \protect\cite[this produces  SRG /eROSITA X-ray bump predicted by][]{2019MNRAS.487.5618L}. At later stages the  high energy emission is generated in a PWN-manner, at the termination shock of highly relativistic and highly magnetized pulsar wind.} 
\label{outline}
\end {figure}

Small ejecta mass is the key ingredient of a model by  \cite{2019MNRAS.487.5618L}, whereby  FBOTs result from 
 an electron-capture  collapse to a \NS\  of a merger product   a massive ONeMg white dwarf (WD) with  another WD, Fig. \ref{outline}.  Two distinct evolutionary channels  lead to the disruption   of the less massive  WD  during the merger and the formation of  a shell  burning non-degenerate star  incorporating  the ONeMg core.    After the transients settle down, the  result is  a special  type shell-burning star with a  size few times $10^9$ cm,  fast rotating (at the surface), with luminosity $L\sim 10^4 L_\odot$, producing nearly hydrogen-clear  winds.     The star lives for $\sim  10^4$ years, while the  envelope mass is  both lost to the wind and added to the core as nuclear ashes. 
If the mass of the core exceeds the Chandrasekhar mass, the
electron-capture collapse follows  after  $\sim 10^2-10^4$ years.

The collapse produces various observed phenomena that depend on the particular properties/parameters of the merging system. In particular,
the observed properties of the collapse depend  on (i) duration of shell burning affecting the amount of envelope mass left at the moment of collapse; (ii) duration of shell burning  and the corresponding amount of  angular moment transferred to the core; (iii)   the viewing angle with respect to the axis. Eventually,  the amounts of mass left in the  shell and the core's  angular momentum at the moment of collapse depend on the masses of the merging WDs and  the orbital separation of the Main Sequence stars.
This scenario explains   a small envelope mass of FBOTs:  as little as $\sim 10^{-2} M_\odot $ of the material is ejected with the total energy $\sim$ few $ 10^{50}$ ergs. This ejecta becomes optically thin on a time scale of days.

During the collapse,  the   neutron star is spun up and magnetic field is amplified  \citep{2009A&A...498..241O,2011MNRAS.415..944B,2015Natur.528..376M}. 
The ensuing fast magnetically-dominated relativistic wind from the newly formed neutron star shocks against the ejecta, and later against the pre-collapse  wind.
The radiation-dominated forward shock produces the long-lasting optical afterglow, while the  termination  shock of the  relativistic  wind produces the  high energy emission   in a manner similar to Pulsar Wind Nebulae. 

\begin{figure} 
\includegraphics[width=.49\linewidth]{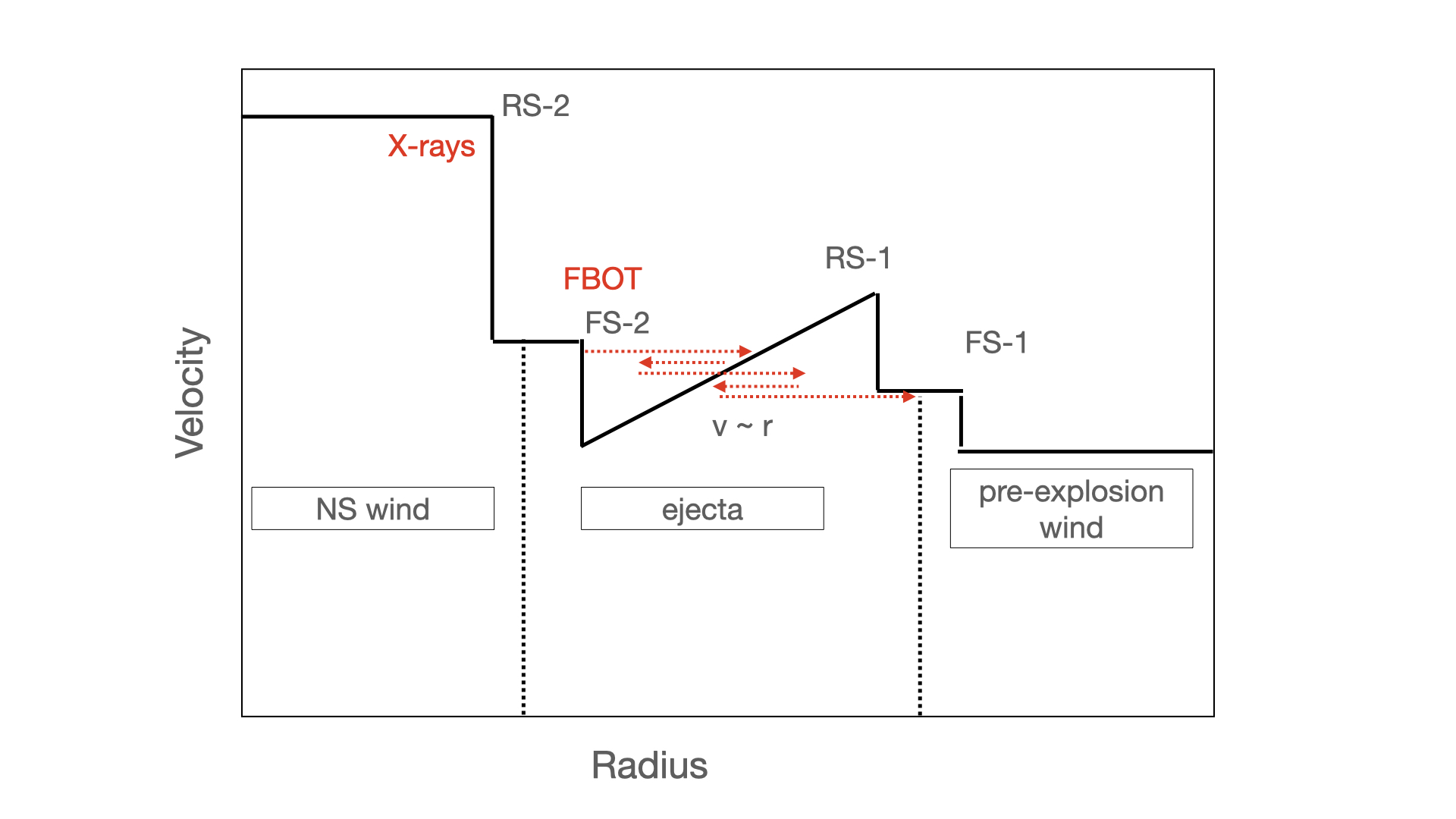}
\includegraphics[width=.49\linewidth]{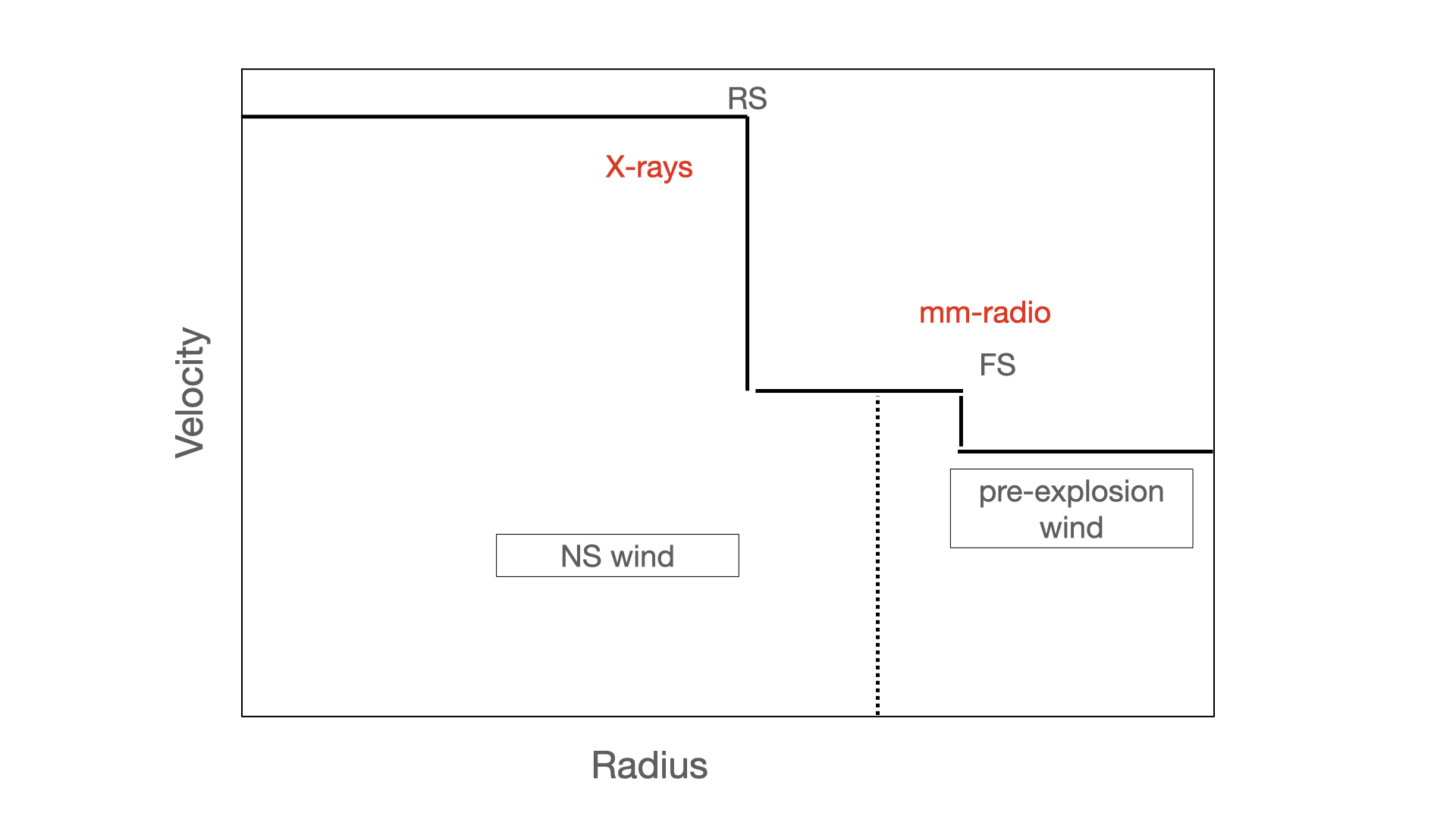}
\caption{Velocity structure: Left panel before NS-drive shock  breaks out into the wind (top right panel in Fig. \protect\ref{outline}); Right panel after break-out  (bottom right panel in Fig. \protect\ref{outline}). At large radii this is a  pre-explosion wind with  $v_w \sim 10^4$ km s$^{-1}$. Ejecta, with a  linear velocity profile, propagates into the wind, launching a forward shock (FS-1) and reverse  shock (RS-1). At smaller radii the wind from the \NS\ generate forward shock (FS-2)  in the ejecta and reverse  shock (RS-2) in the \NS\ wind. X-rays are produced at  RS-2, FBOT  proper at FS-2. Photons (red dashed lines) diffuse ahead of the radiation-dominated shock. At later times  the \NS\ wind may broke out into the pre-explosion wind. Long term X-rays  are generated at the   RS, mm-radio emission  - at the FS.} 
\label{veclocity}
\end {figure}

\section{FBOT proper}

In the case of AT2020mrf most constraints come from the late X-ray detection. In this section we address the nature of FBOT proper (bright optical transients lasting a few days), but we use numerical parameters demanded by the X-ray, see \S \ref{Xray}. 

\subsection{Small ejecta mass}
\label{small}

An  AIC of a WD  produces a central engine,  a fast rotating \NS, while ejecting some mass  $M_{ej}$. 
For homologous expansion of the ejecta with $ v \propto r$ \citep[see][for a more detailed modeling]{1956JFM.....1..436S,1999ApJ...510..379M,2013ApJ...773...79R}, the energy in the ejected part is
\be
E_{ej} = \frac{3}{10} M_{ej}V_{ej}^2,
\ee
while the   density evolves according to 
\be
\rho_{ej} = \frac{3}{4 \pi} \frac{M_{ej}}{(V_{ej} t_{ej} )^3} 
\label{rhoej}
\ee
where $V_{ej}$ is the maximum velocity of the ejecta. 

Using scattering cross-section $\kappa \approx 0.1 $ cm$^2$ g$^{-1}$ \citep{1982ApJ...253..785A}, total optical depth through ejecta 
\be
\tau _{tot} = \frac{3}{ 4 \pi} \frac {M_{ej} \kappa}{ V_{ej} ^2 t^2} = 640   m_{ej,-1} t_d^{-2}  V_{ej,4} ^{-2}
\label{tautot}
\ee
where  $m_{ej,-1} =  M_{ej}/(0.1 M_\odot)$, $ V_{ej,4} =  V_{ej}/(10^4 {\rm km s} ^{-1})$ and    time $t_d$ is measured in days.

The diffusion time is then 
\be
t_{FBOT} = \left( \frac{3}{4\pi}  \frac {M_{ej} \kappa  } {  c V_{ej}} \right)^{1/2} = 4.6 \, m_{ej,-1}^{ 1/2}\,   V_{ej,4} ^{-1/2}  \,  {\rm days} 
\label{tFBOT}
\ee
We identify the diffusion time $t_{FBOT} $ with the observed  peak of the light curve. 

Several arguments can be used to derive (\ref{tFBOT}). In \S \ref{GreenC} we derive photon's Green function in  the expanding ejecta, showing the  scaling  for the diffusion radius $r_d \propto t^2$, Eq. (\ref{rdiff}).  Second, the  time for a shock traveling with velocity $V_s$ through ejecta of thickness $V_{ej} t$ should be set equal to the photon  diffusion time $ (V_{ej} t)^2/( c l_{mfp} )$ where $ l_{mfp}= 1/(\kappa \rho)$ is mean free path \citep{1963PThPh..30..170O,1972ApJ...178..779C}; equating the two gives  (\ref{tFBOT}). Equivalently, 
for radiation-dominated shocks,  the photons diffusing ahead of the shock escape when optical depth to the emitting surface is  $\tau \sim c/V_s$.
Estimating  shock velocity $V_s \sim  V_{ej}$ this occurs at (\ref{tFBOT})  \citep[see also][]{1982ApJ...253..785A}. 

\begin{figure} 
\includegraphics[width=.99\linewidth]{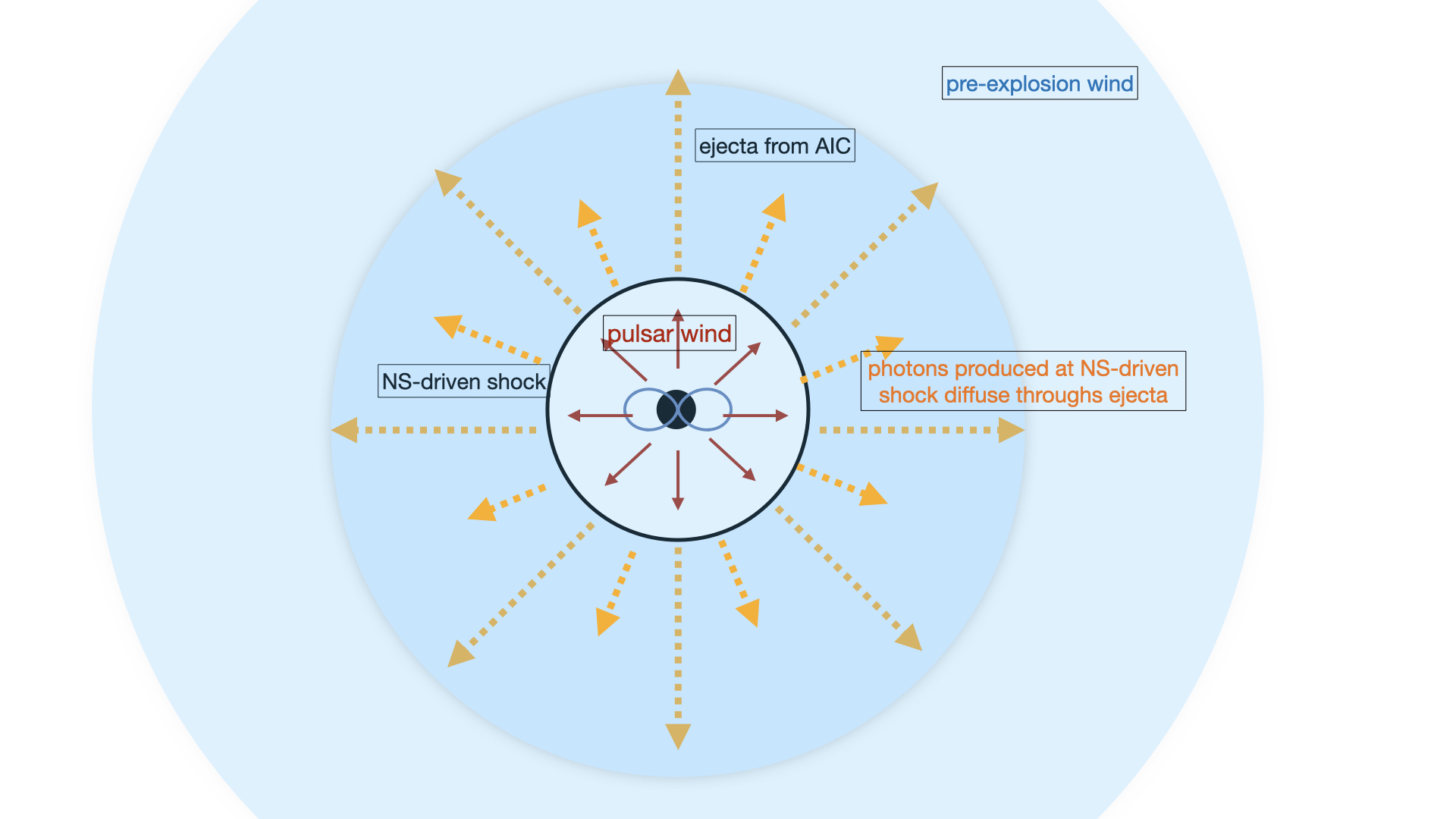}
\caption{Diffusion of NS-shock produced photons in optically thick ejecta. The diffusion time corresponds  approximately to  the rise time of FBOTs, Eqns. (\ref{tFBOT}) and   (\ref{tFBOT1}) } 
\label{photons-diffuse}
\end {figure}

If scattering is purely Thomson, then one needs to change  $\kappa \to \sigma _T /m_p = 0.4$ cm$^2$ g$^{-1}$,
\ba &&
\tau _{tot} = \frac{3}{ 4 \pi} \frac {M_{ej} \sigma_T}{  m_pV_{ej} ^2 t^2} = 2.5 \times 10^3 \times   m_{ej,-1} t_d^{-2}  V_{ej,4} ^{-2}
\nn &&
t_{FBOT} = \left( \frac{3}{4\pi}  \frac {M_{ej}\sigma_T  } {  c  m_p V_{ej}} \right)^{1/2} = 9.2\, m_{ej,-1}^{ 1/2}\,   V_{ej,4} ^{-1/2}  \,  {\rm days} ,
\label{tFBOT1}
\ea
 a mild correction. The gray scattering cross-section $\kappa \sim Y_e x_e$ cm$^2$ s$^{-1}$, $Y_e $ is the electron fraction, $x_e$ is the ionization degree.

For AT2020mrf the peak is at  3.8 days, hence
\ba && 
 M_{ej} = 6.7 \times 10^{-2} M_\odot V_{ej,4} ^{1/2} 
 \nn &&
 E_{ej} = \frac{3}{10} M_{ej}  V_{ej}^2 = 4 \times 10^{49}  V_{ej,4} ^{2}   {\rm erg}
  \nn &&
 R_{ej} =  V_{ej} t_{FBOT}  = 3 \times 10^{14} {\rm cm}
 \label{RR} 
 \ea
 (At times much shorter than (\ref{tau23}) the photospheric radius  is just  a bit smaller, by $\sim 1- (2/3)  V_{ej} /c$.) 

Estimate  (\ref{RR}) implies that ejecta mass must be small, $\leq 10^{-1} M_\odot$, much smaller of a typical  ejecta of any conventional supernova.
 If  $\sim 1 M_\odot$ is ejected, short diffusion time (\ref{tFBOT}) would require $V_{ej} \sim c$, with the total energetics at $\sim 10^{54} $ ergs. Estimate (\ref{RR}) on the mass is the upper limit: (i)   \cite{2015MNRAS.450.1295W}  estimated (\ref{tFBOT1}) as the rise time (not peak time), which is shorter; (ii) we omitted a factor $1/3$ in the diffusion coefficient.

 The ejecta itself becomes  fully transparent at somewhat longer times
\be
t_{\tau} \approx t_{FBOT} \sqrt{ \frac{c}{V_{ej}} } = 25\, m_{ej,-1}^{ 1/2}\,   V_{ej,4} ^{-1}  \,  {\rm days}  \to  21   V_{ej,4} ^{-1}  \,  {\rm days}
\label{tau23} 
\ee
(the last relation uses estimate of the ejecta mass (\ref{RR})). 
Thus the FBOT proper is contributed both by the  energy input from the NS wind and somewhat later by cooling of the ejecta.

The effective emission radius starts to decrease at approximately half of the  time (\ref{tau23}). 
Qualitatively, the radius $R_{ph}$ where  optical depth to the surface equals unity evolves according to
\be
 R_{ph} = V_{ej} t \left( 1 - \frac{4 \pi}{3} \frac{(V_{ej} t )^2} {2 M_{ej} \kappa} \right)
\ee
It  decreases after  half of the  time (\ref{tau23}),  until the ejecta  becomes  fully transparent  at time (\ref{tau23}).
 
This is consistent with observation of \cite[][their Fig. 8]{2019MNRAS.484.1031P} which shows long-term decreasing effective radius (ejecta velocity somewhat smaller that $10^4 $ km s$^{-1}$ is required to extend this to $\sim 1 $ month). We also note that similar to early stages of terrestrial nuclear explosions, the radiative cooling of the ejecta may be faster via the formation of the cooling wave \cite[][]{ZeldovichRaizer,1976Ap&SS..44..429G}.

Small mass of the ejecta is a challenge to the conventional  models: typical SN  ejected mass is $\geq 1 M_\odot$. To produce  mildly relativistic motion  of  large mass, even for jetted outflows, requires GRB-type energies, well  in excess of $10^{51}$  of kinetic energy (and correspondingly large total rates.)   The model of  \cite{2019MNRAS.487.5618L} offers a natural explanation. 

\subsection{Blue color: radiation-dominated shock within the ejecta}

Consider a two-stage explosion: the initial neutrino-driven ejection of a shell with mass $M_{ej}$ and maximal velocity $V_{ej}$, followed by the central source (\NS-)  driven (second) wind
(the ``first wind'' is the pre-explosion wind from the progenitor star.)  Thus, the initial ``heavy lifting'' is done by SN shock. The energy of the central engine is then mostly spent on producing the non-thermal emission, as opposed to generating heavy slow outflows \citep[see discussion by][in applications to GRBs]{2011MNRAS.411.2054L}. The required delay time is only few seconds - time for the SN shock to cross the collapsing envelope. In the magnetar model of relativistic explosion  \citep{1992Natur.357..472U,2007MNRAS.382.1029K,2011MNRAS.413.2031M}  it is expected that in few seconds the \NS\ cools sufficiently so that it's wind becomes clean, pulsar-like.

After the SN shock propagated through the ejecta-to-be, 
the NS-driven wind launches  another shock in the expanding envelope. We consider its dynamics next. 

Qualitatively two regimes for the shock propagation through ejecta may be realized: energy conserving Sedov-type \citep{Sedov},  and momentum driven (thin shell) Kompaneets-type \citep{1960SPhD....5...46K}. Sedov-type flow would occur if the termination shock in the wind is close to the source (\eg as in the case of Crab PWNe). The wind-blown bubble is then in approximate causal contact. The thin shell case is realized if the termination shock is close to the contact discontinuity -  it is much more energetically demanding, by a factor $c/V_s$, and is less likely to be realized.  Below we use the Sedov scaling \citep[see also][]{Chevalier82}

At sufficiently short times, less than a  month, we can assume a constant spin-down luminosity $L_{sd}$ (see Eq. (\ref{tOmega}) for justification).  For a contact discontinuity at radius $R$ the swept-up energy and mass are
\ba &&
E_s = \frac{3}{10} \frac{  M_{ej}  R^5}{V_{ej}^3 t^5}
\nn &&
M_s = \frac{  M_{ej}  R^3}{V_{ej}^3 t^3}
\ea
(since $R \propto t^{6/5}$, Eq. (\ref{rrr}), both $E_s$ and $M_s$ increase with time).

The energy balance and the equation of  motion then read
\ba &&
L_{sd} t + E_s = \frac{ M_s (\partial_t R)^2}{2}
\nn && 
\partial_t R = \sqrt{  \frac{2 L_{sd} V_{ej}^3 t^4 }{M_{ej} R^3 }  + \frac{3}{5} \frac {R^2}{t^2}} 
\label{Sedov}
\ea

Eq. (\ref{Sedov}) has a solution 
\ba && 
R= \left(  \frac{50}{21} \frac{L_{sd} V_{ej}^3 t^6}{M_{ej}}\right)^{1/5} 
\nn && 
\frac{R}{V_{ej} t} =  \left( \frac{t}{t_{br}}  \right) ^{1/5}
\label{rrr}
\ea
Velocity of the shock with  respect to the ejecta
\be
V_s = \partial_t R - \frac{R}{t}  = \left( \frac{t}{t_{br}}  \right) ^{1/5}  \frac{V_{ej}}{5}
\ee
where $t_{br}$ is given by  (\ref{tbr}).

The radiation-dominated shock  jump condition (it may be verified that early on the shock is radiation-dominated) 
\be
\frac{4}{c} \sigma_{SB} T^4 \sim \rho_{ej} V_s^2
\ee
gives
\be
T = 1.5 \times 10^3 \frac{ L_X^{1/10} M_{ej} ^{3/20}  }{ t^{13/20} V_{ej} ^{9/20} \epsilon_X^{1/10} } = 4 \times 10^4 
L_{X,42} ^{1/10} \epsilon_{X,-1} ^{-1/10}  V_{ej,4} ^{-9/20}  \, K
\label{TT}
\ee
at time $t= 3.8$ days. 
This is an upper limit on temperature as it assumes that all the enthalpy is provided by the radiation; matter contribution would reduce the estimate of temperature. 
A blackbody fit to the spectrum suggests a temperature of $T = 2  \times  10^4$ K and a radius of  $R = 7.9  \times  10^{14}$  cm  \citep{2021arXiv211200751Y}. Our estimates (\ref{RR}) and (\ref{TT}) are consistent.

Predicted evolution of temperature $T(t) \propto t^{-13/20} $ (\ref{TT}),  corresponding  to the constant density of the ejecta, can be tested against observations. Our fit to the evolution of temperature in FBOT AT2018cow \citep[Table 4 of][]{2019MNRAS.484.1031P} gives $T \propto t^{-0.23}$  \citep[limited to 27 days;  after this time  there is a clear break in the evolution of temperature; in our model the break is due to the transition of the NS-powered shock into the preceding wind][]{2019MNRAS.487.5618L}.
Freely expanding ejecta models predict much steeper decay  $T \propto t^{-1}$ (in the adiabatic limit). Thus the present  model fairs better. Also, variations of the density of the ejecta give more freedom. 

Additional complication comes from the fact that  in the highly radiation-dominated regime the shock, defined as a hydrodynamic discontinuity, may disappears, so that fluid properties evolve smoothly  \citep{ZeldovichRaizer,1976ApJS...32..233W}. Alternatively,  an isothermal jump may form \citep{LLVI}. Qualitatively, the above estimates remain valid.

The energetics of FBOTs comes both from the internal energy of the hot ejecta and the luminosity of the central source. Both sources of energy also experience adiabatic losses.  For the energy contributed by the  central source all the photons emitted by the central source at times $t \leq t_{FBOT}$ come out in a narrow window $\sim 1/3$ near $t_{FBOT}$,  see Fig. \ref{FBOT-MonteCarlo}).  In addition,  a photon that is injected at day 1 and finally diffuses out on day 3 will have dropped in energy by a factor of $\sim 1/3$.  
  We can estimate the peak luminosity  then as $ L _{FBOT} \sim   L_{sd} \sim 10^{44}$ erg s$^{-1}$.


\subsection{Short duration of FBOTs: photon diffusion from the central source through  expanding ejecta}
\label{GreenC}

\subsubsection{Photon's Green's function in expanding ejecta}
\label{Green}

\begin{figure} 
\includegraphics[width=.99\linewidth]{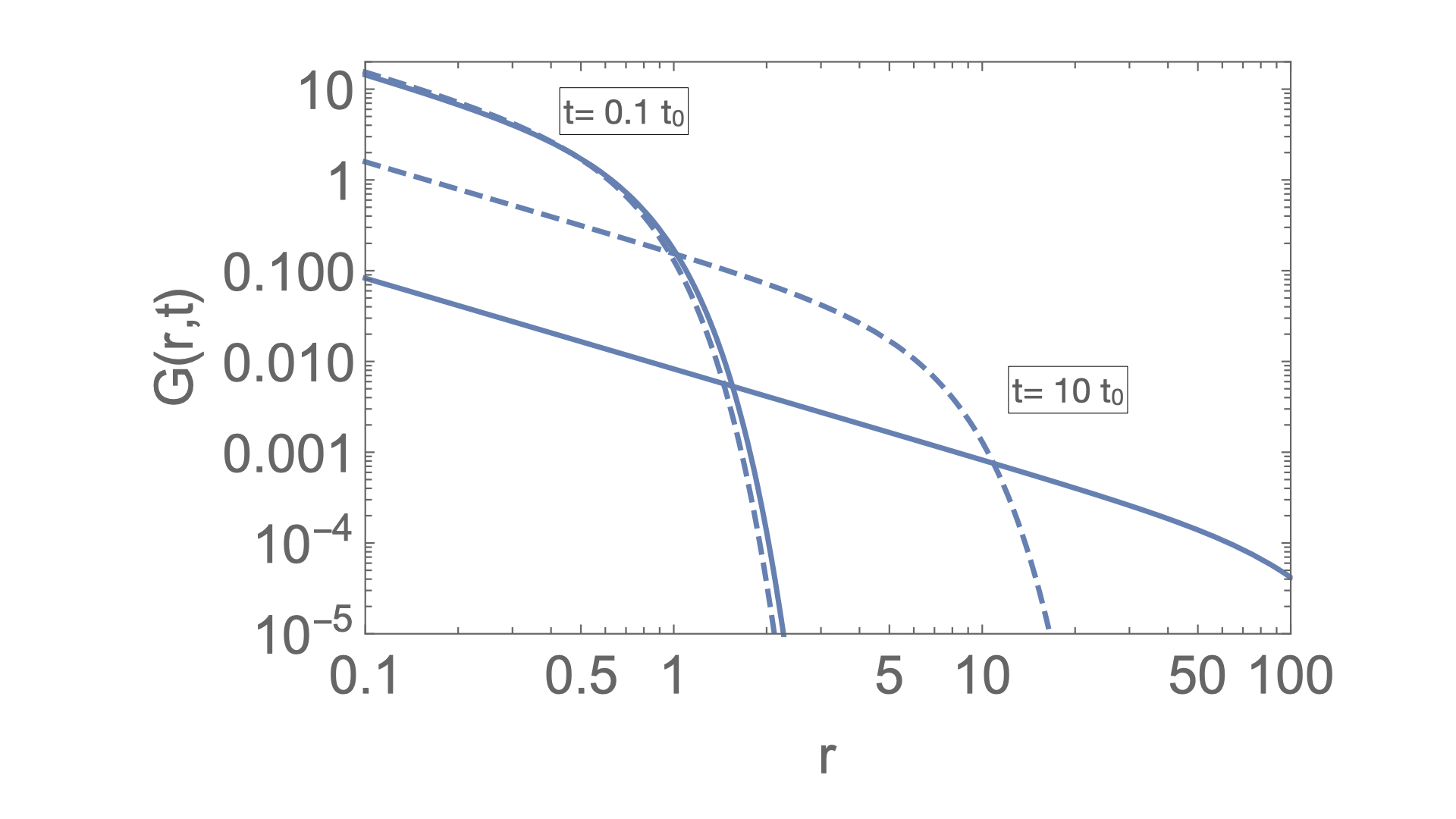}
\caption{Photons' Green function for diffusion in expanding ejecta  with time-dependent diffusion coefficient (\protect\ref{Green1}) (solid lines) if compared with Green's function for constant diffusion coefficient (dashed lines).  Two sets of plotted  curves correspond to $t=0.1 t_0$ and $t=10 t_0$.  Notice that at $t=10 t_0$ the Green's function extends to $\sim  t^2$.  } 
\label{photonGreen}
\end {figure}

Consider early times when the NS-powered  shock is at small radii within the ejecta. Approximate the shock as a source of photons located at $r=0$ (according to (\ref{rrr}) the shock remains deep with the the ejecta for a long time).  In an expanding   medium the  diffusion coefficient $\eta_d$  varies as
\ba &&
   \eta_d=\kappa_0 ( t_++t_0)^3
   \nn &&
\kappa_0 = \frac{4 \pi}{3}  \frac{ c V_{ej}^3}{M_{ej} \kappa},
\ea
where $t_0$ is the delay time  before the beginning of the expansion and the injection of photons and $ t_+$ is time since injection. Neglecting advective  transport, 
  making a change in time variable, 
 \ba && 
 \tilde{t} = \frac{1}{4} t \left( t_+^3+4 t_0  t_+^2+6 t_0^2  t_++4 t_0^3\right)
 \nn && 
 \tilde{t} \approx t^4/4\, \mbox{for} \,  t \gg t_0,
 \ea
 the diffusion equation reduces to the familiar form with solution
\be
G(r,t) = 
\frac{ 1 }{8 (\pi {\tilde{t}  \kappa_0})^{3/2} }  e^{- r^2/(4 \kappa_0 \tilde{t} )}
\label{Green1} 
\ee
see Fig. \ref{photonGreen}. $G(r,t)$ is normalized so that  $ 4\pi \int _0 ^\infty r^2 G(r,t) dr =1$.
Eq. (\ref{Green1}) gives Green's function for $\delta(t)$ injection at $r=0$ that occurred   at $t_0$ after  explosion, as measure at time 
$t_+$ after the injection. 

Shifting time $ t_+ \to  t - t_0$ ($t $ is time measured from the initial explosion),
\be
G(r,t) = 
\frac{ 1 }{8 (\pi (t^4-t_0^4)  \kappa_0)^{3/2} }  e^{- r^2/( \kappa_0 (t^4-t_0^4) )}
\label{Green1} 
\ee
This is Green' s function for injection at time $t_0$ after the explosion, as measured at time  $t$ after the explosion; $ t > t_0$.

For zero delay, $t_0=0$,
\be
G(r,t) = 
\frac{ 1 }{8 (\pi   \kappa_0)^{3/2} t^ 6 }  e^{- r^2/( \kappa_0 t^4 )}
\label{Green2} 
\ee
 
 Thus,  the diffusion   radius scales as
\be
r_d \approx  \sqrt{\kappa_0} t^2
\label{rdiff}
\ee
The  diffusion  radius (\ref{rdiff})  then becomes $\sim V_{ej} t$ at time  (\ref{tFBOT})-(\ref{tFBOT1}).

\subsubsection{Short duration: ``diffusive caustic''}

During the FBOT proper the wind shock remains deep inside the ejecta. We can then use the Green's function (\ref{Green1}) with a given wind luminosity to find distribution of photons inside the ejecta. (One also  has to take account of the fact that photons escape from the edge.) Let us give semi-qualitative estimates of the expected light curve.

\begin{figure} 
\includegraphics[width=.99\linewidth]{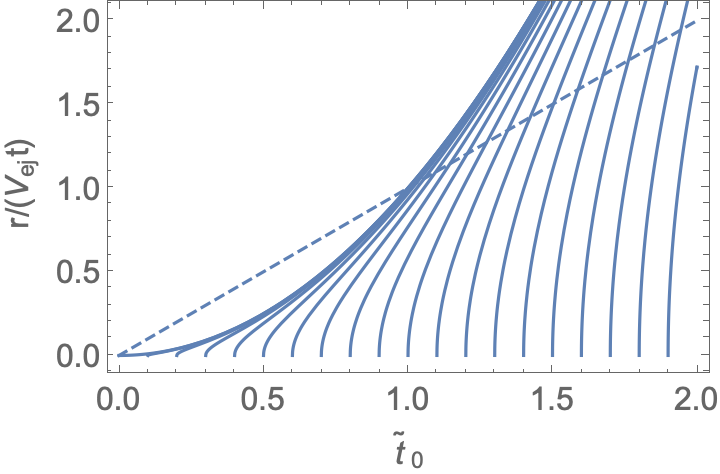}
\caption{ Photon trajectories  (\ref{rdd}) as a function of the injection time $ \tilde{t}_{0}$. Dashed line with inclination of 1 is the surface of the ejecta. In this approximation nearly all the photons produced before $t_{FBOT}$ escape nearly simultaneously. } 
\label{tescoft0}
\end {figure}

Using (\ref{Green1}) the typical photon trajectory is 
\be
r_d = \sqrt{ t^4-t_0^4} \sqrt{\kappa_0}
\label{rdd} 
\ee
(Again, here $t_0$ is injection time, $t$ is time since the explosions). Equating $r_d$ with the location of the surface of the ejecta,   $r_d = V_{ej} t$, we find the escape time

\ba &&
\tilde{t}_{esc}  = \frac{\sqrt{1+ \sqrt{4 \tilde {t}_0^4+1}}}{\sqrt{2}}
\nn && 
\tilde{t}_{esc} =\frac{{t}_{esc} }{t_{FBOT}}
\nn && 
\tilde{t}_{0} =\frac{{t}_{0} }{t_{FBOT}}
\ea
Thus, all the photons produced by the central source deep inside the ejecta, $ \tilde {t}_0 \leq 1$ escape almost simultaneously, Fig. \ref{tescoft0}. This reminds of a caustic, hence the name.

To get an analytical estimate of the flux, we can integrate (\ref{Green1}) times the luminosity over the injection time $t_0$. For constant photon production rate, using steepest decent method we find the photon density at $r = V_{et} t$
\be
n_{ph} (r= V_{et} t) \propto \frac{1}{t_{FBOT} }  \frac{1}{t^4 }\frac{1}{\left(1-(2/3)  (t_{FBOT} /t)^2\right)^{3/4} }
\label{nph}
\ee
Divergence at $t= \sqrt{2/3} t_{FBOT}$ is an artifact of the analytical approximation. Yet it demonstrates qualitatively  that short, bright events can be produced. 

Relation (\ref{nph}) also explains why short  $t_{FBOT}$ are needed to produce bright FBOTs: the flux at $t=t_{FBOT}$  is  (neglecting the divergent component) 
\be
F  \propto  \left. n_{ph}\right|_{r = V_{et} } \times r^2 \propto t_{FBOT} ^{-3}
\ee

\subsubsection{Monte Carlo simulations of photon escape}

\begin{figure} 
\includegraphics[width=.49\linewidth]{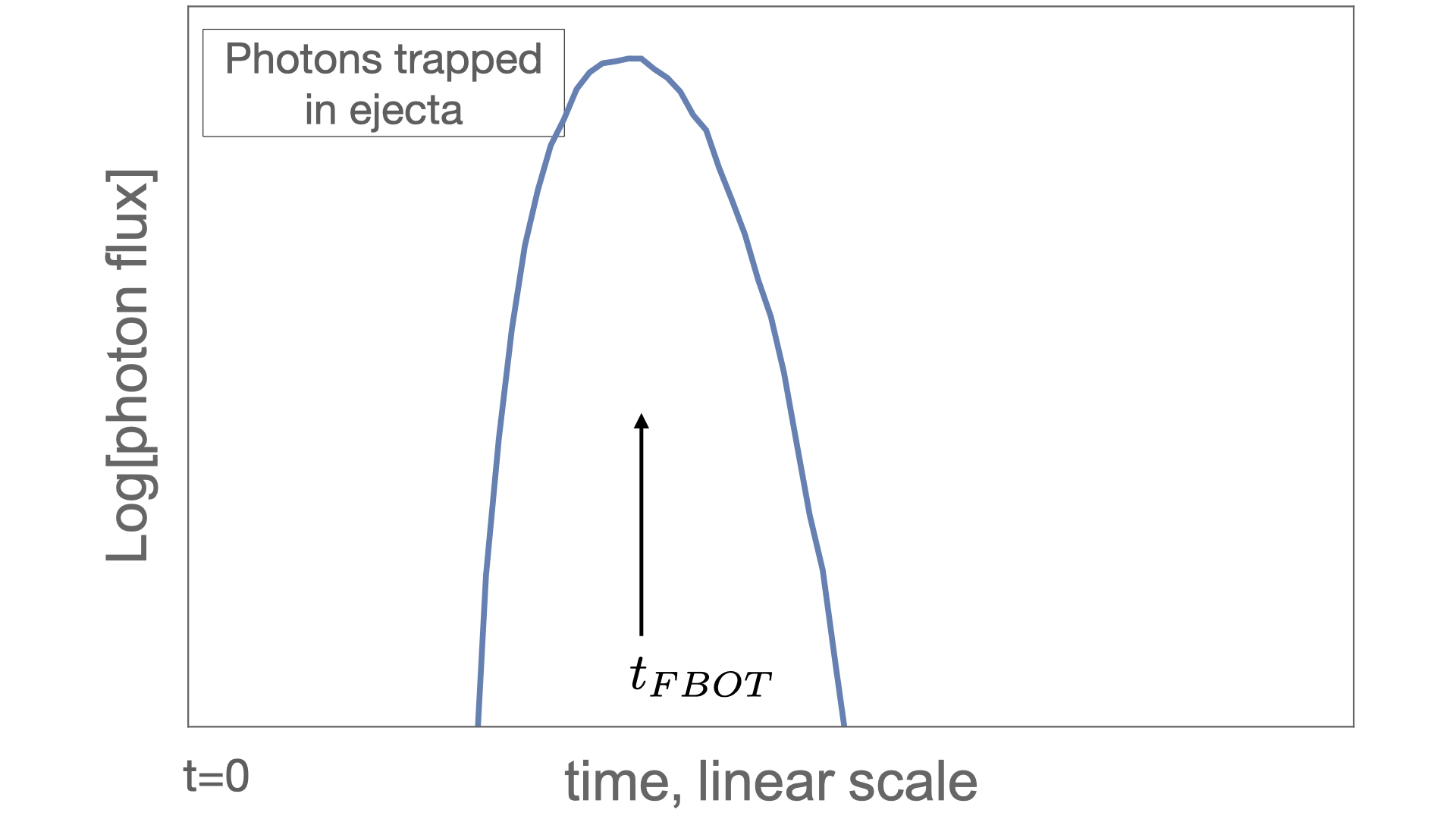}
\includegraphics[width=.49\linewidth]{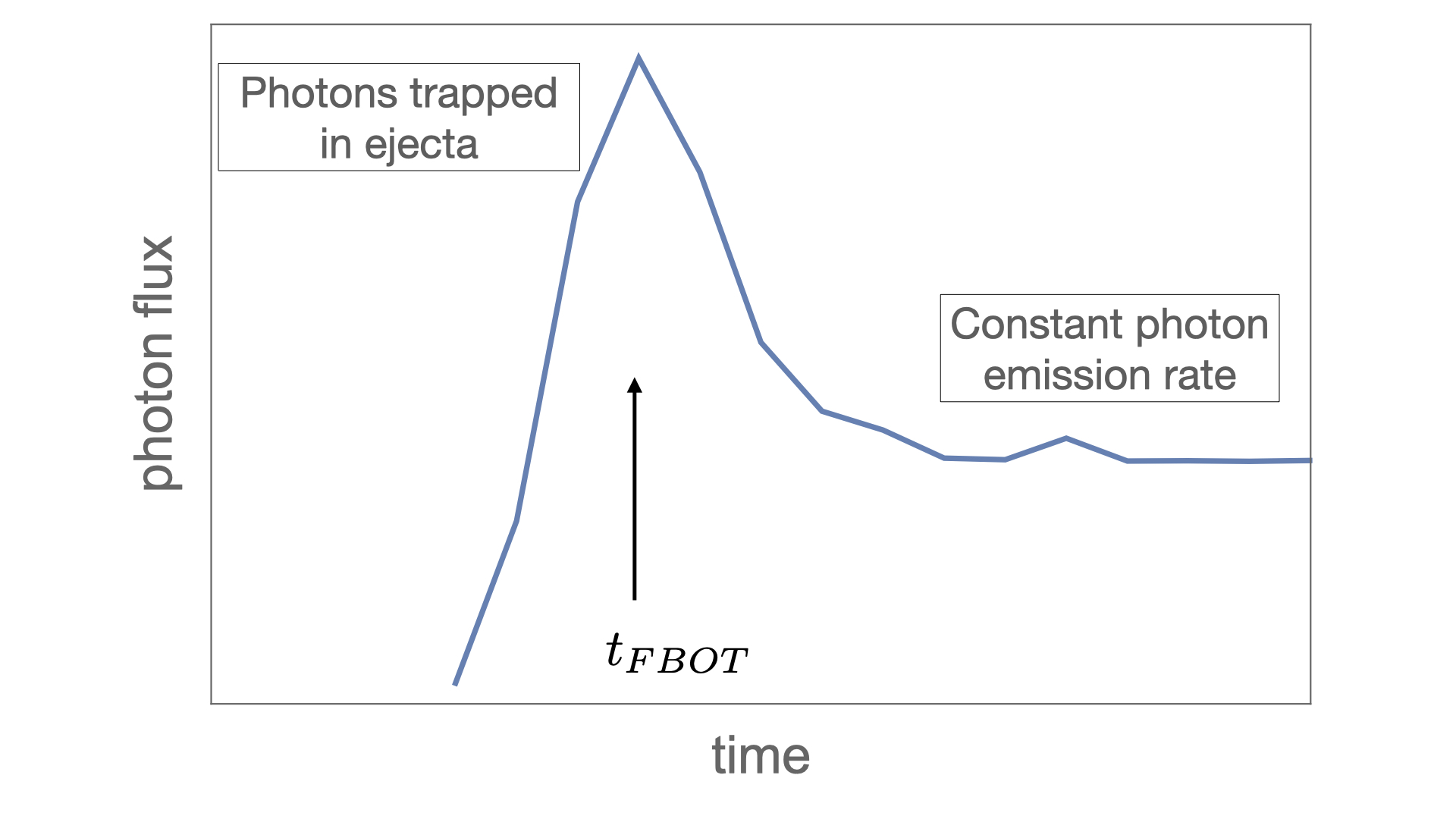}
\\
\includegraphics[width=.49\linewidth]{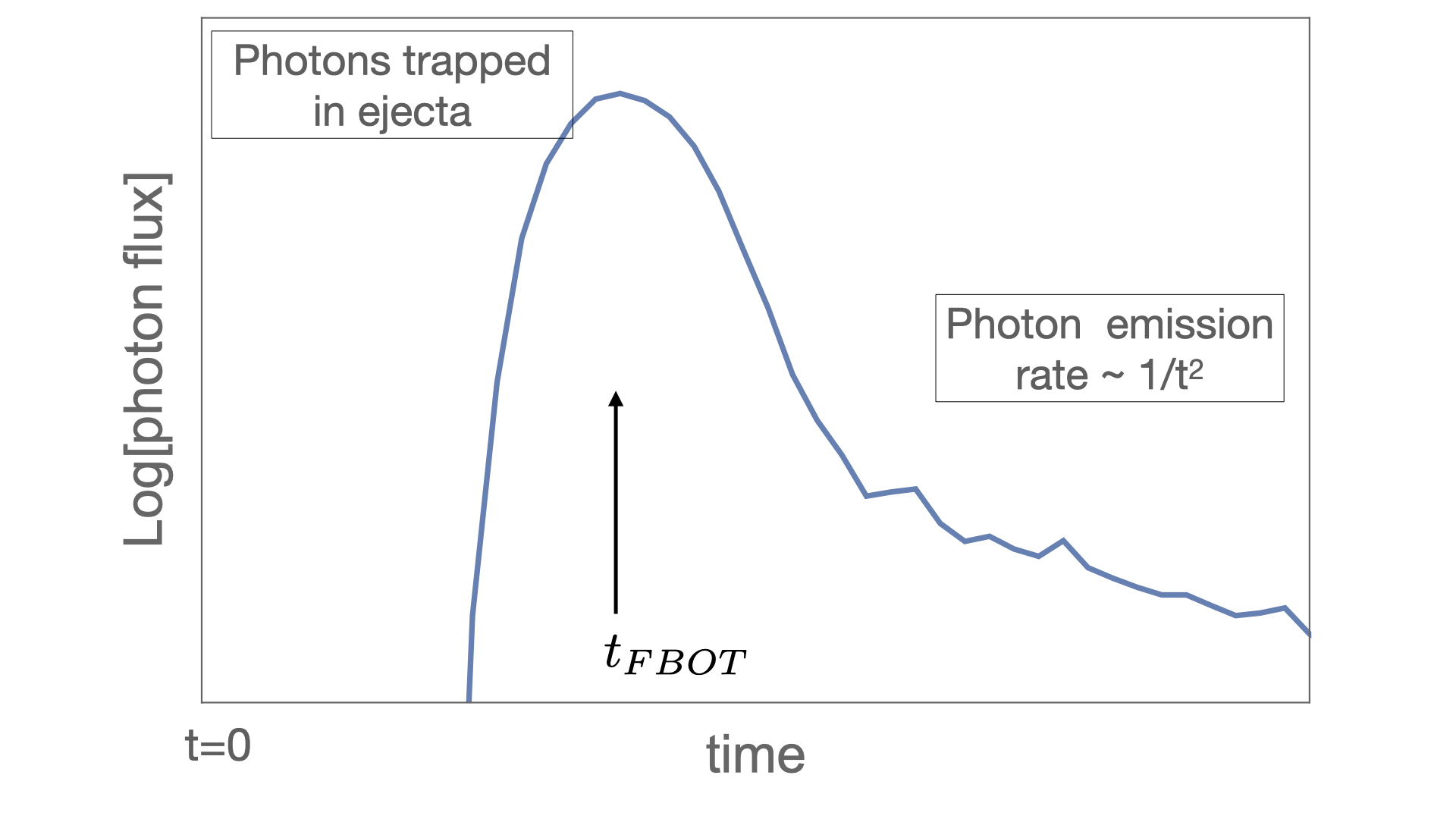}
\includegraphics[width=.4\linewidth]{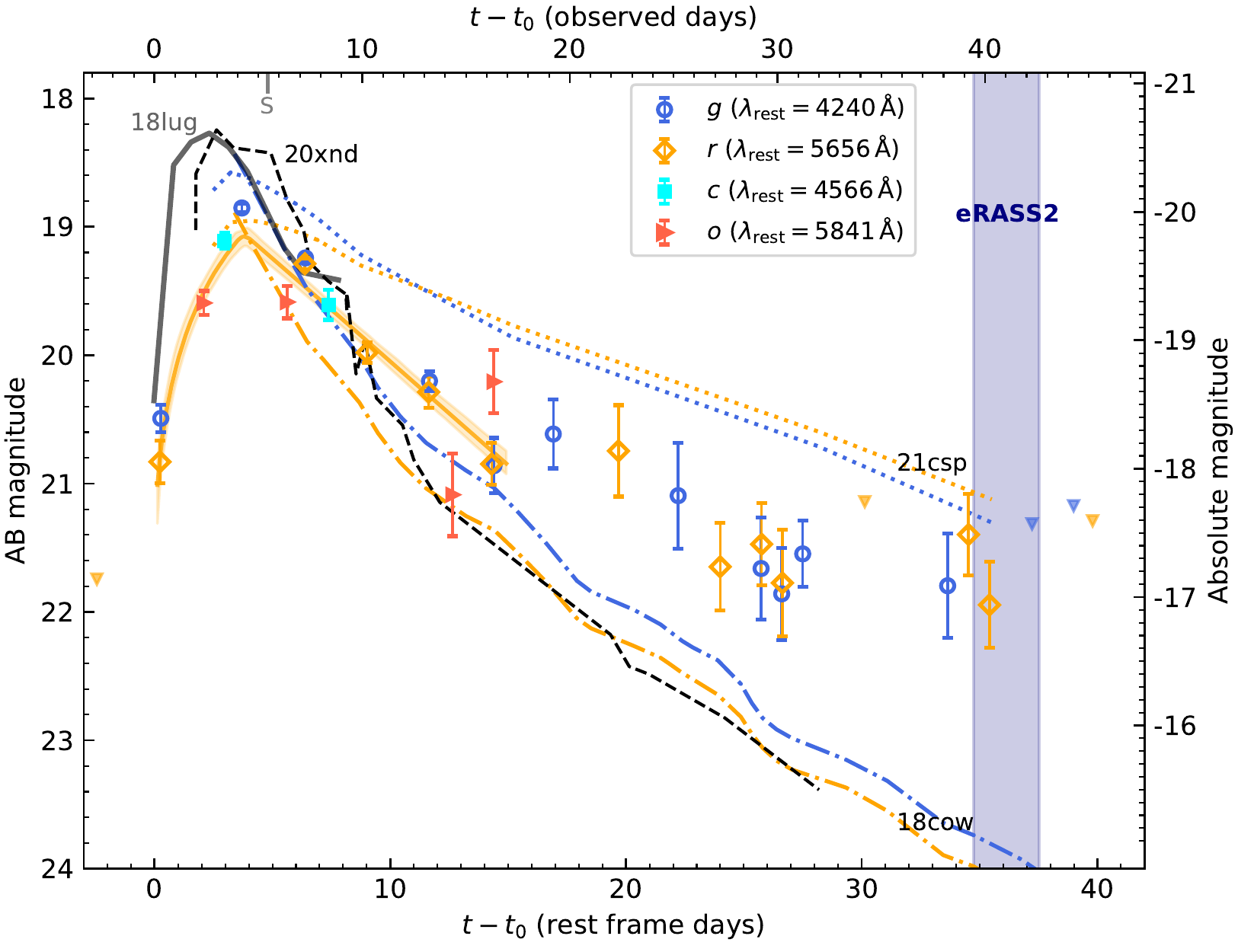}
\caption{1D Monte Carlo simulations of photons emitted by the NS-driven shock propagating  within the ejecta. Photons are injected according to various prescriptions for spin-down luminosity. Plotted is the escaping flux. Top left: photons are injected only at time $t_0=0$. This is the expected light curve for an engine with short activity time. Top right: constant luminosity source. 
 Narrow peak near   $t_{FBOT}  $  occurs when, approximately, all the photons emitted by the central source at times $t \leq t_{FBOT}$ escape diffusively. The peak photon flux rate is $\sim 3 $ times the average photon emission rate corresponding to the constant flux at later times (in this panel the flux  scale is linear).  
 Bottom left: Decreasing photon emission rate $\propto (1+\tilde{t}_0)^2$.  The model compares favorably with observations:  the light curve  of   AT2020mrf   \citep{2021arXiv211200751Y},  showing a prominent peak and decaying tail, bottom right  panel.}
\label{FBOT-MonteCarlo}
\end {figure}

We performed simple 1D Monte Carlo simulations of photon propagation within the ejecta. Photons are injected at $r=0$ with variable rate (impulsive $\delta(t)$, constant and declining rates). The diffusion coefficient scales as $\propto t^3$.  Coordinate of each photon  $r$ is traced until it reaches $V_{ej} t$. It is then assumed that the photon scapes. In Fig. 
\ref{FBOT-MonteCarlo} we plot the rate of photon escape. At early times all the photons are trapped.  Near $t_{FBOT}  $ all the  trapped photons  escape with a short time $\sim t_{FBOT}/3$.  Thus, the peak flux is $\sim $ three times above the average source luminosity. These are  clear counter example to the  ``Arnett's law''  \cite[][that the luminosity at the peak is equal to the instantaneous luminosity  at that time]{1982ApJ...253..785A}. For example, for impulsive injection the instantaneous luminosity at the peak escaping luminosity is zero. 
 At later times the flux is approximately the injected flux (zero for  impulsive, constant and/or declining).

Note that  at long times the optical light curve reflects the (decreasing)  rate of the production of the optical photons in the forward shock, not the power of the central engine. which may remain constant. 

Our results for light curves  are different from \cite{2010ApJ...717..245K}.  We treat the input from a magnetar as perturbation, not  as a dominant energy source in the ejecta.

\subsection{SRG /eROSITA X-ray bump: break-out of the NS-driven  shock from ejecta into the wind}
The NS-driven shock breaks from the ejecta at 
\be
t_{br} = \frac{21}{50}  \frac{M_{ej} V_{ej}^2}{L_{sd}} \approx \frac{E_{ej}}{L_{sd}}
\label{tbr} 
\ee
approximately when the NS-injected energy becomes comparable to the ejecta energy. (At times (\ref{ttbr}) the spin-down power can be assumed constant, see Eq. (\ref{tOmega})).

Using (\ref{Lsd}) this occurs at 
\be
t_{br} =  \epsilon _e   \frac{E_{ej}}{L_{X}} = 46   \epsilon _{e,-1}   V_{ej,4} ^{2}    L_{X,42}^{-1} \, {\rm days}
\label{ttbr} 
\ee
(the ejecta mass is incorporated into ejecta energy, see Eq. (\ref{tbr})).

We associate this time with the SRG /eROSITA X-ray bump. \citep[Production of UV and soft  X-rays during shock breakout has been discussed by ][]{1988SvAL...14..449I,1992ApJ...393..742E,2000ApJ...532.1132B,2004MNRAS.351..694C}.
The X-ray  feature during the break-out was predicted by  \cite{2019MNRAS.487.5618L}.

\section{Long term X-ray emission from relativistic  termination shock}
 \label{Xray} 
 \subsection{Dynamics of wind-wind interaction}

\begin{figure}
\includegraphics[width=.99\linewidth]{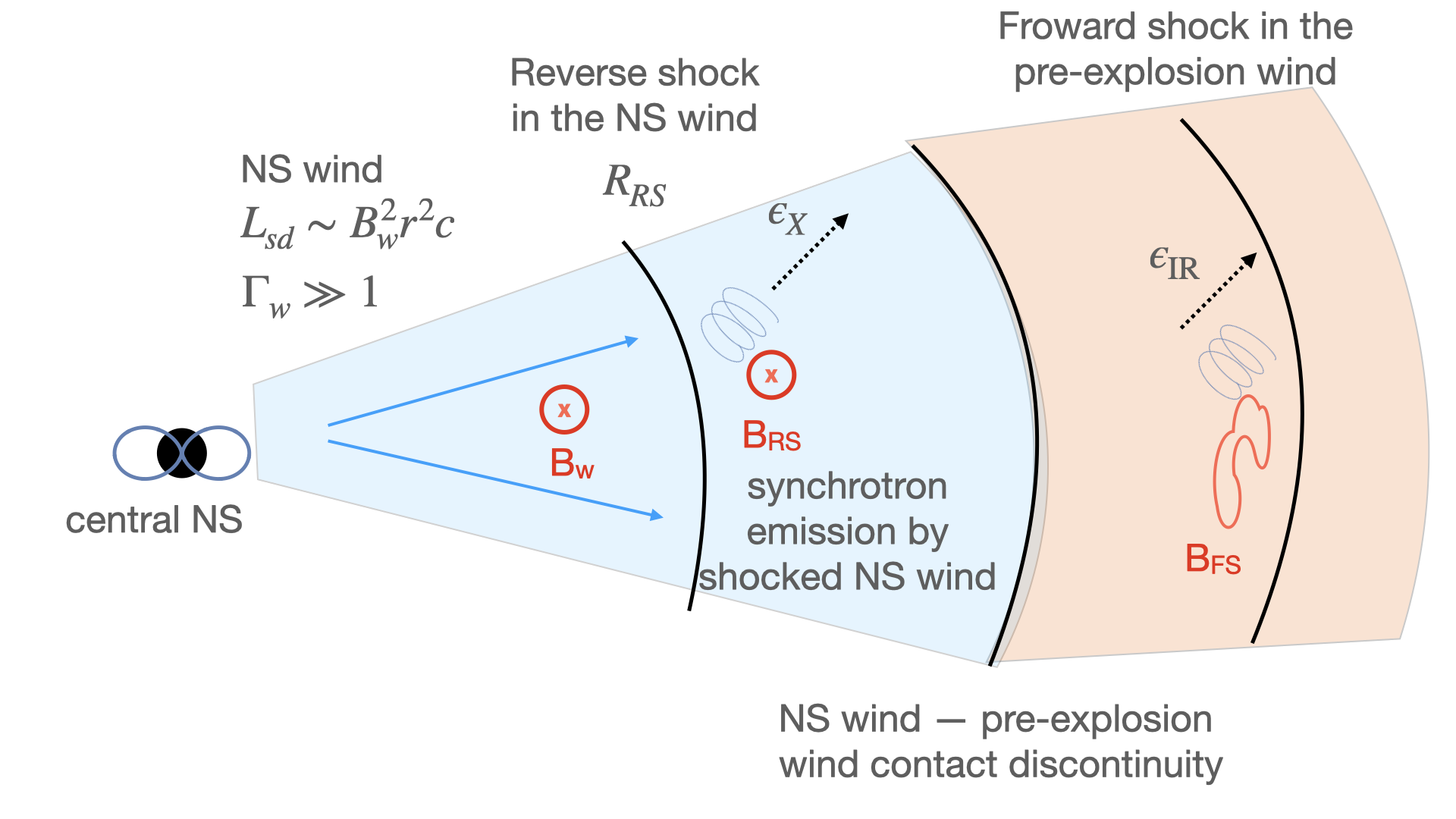}
\caption{Cartoon of late X-ray and IR emission.  The central \NS\ produces a relativistic wind  with \Lf\ $\Gamma_w$ and power $L_{sd}$. The wind shocks at the reverse  shock, where  particles are accelerated to   the  \Lf\ $\gamma \sim \Gamma_w$, and  produce  X-ray synchrotron emission  in the wind's \Bf. 
The IR emission originates from the synchrotron emission  in the  turbulent, amplified  \Bf\ behind the  forward shock.}
\label{FBOTs-New-RS-X}
\end {figure}

The long term X-ray emission provides the most energetic  constraints on the model. At this stage (last panel in Fig. \ref{outline}), the central engine - newly formed \NS\ -  produces  relativistic, highly magnetized wind that produce two shocks, the  forward shock  (FS)  in the preceding wind, and the reverse shock  (RS) in the  relativistic  wind. Thus, it is  relativistic   wind-wind interaction \citep[see][for dynamics of such double explosions]{2017PhFl...29d7101L,2017ApJ...835..206L,,2021ApJ...907..109B}.  Next we discuss the model of \cite{2019MNRAS.487.5618L}  in the light of new observations of  AT2020mrf  by 
\cite{2021arXiv211200751Y}, see Fig. \ref{FBOTs-New-RS-X}  for a qualitative description. 

 At late times, months, the  forward shock  propagates though a powerful pre-explosion wind  with mass loss rate 
\be
\dot{M} = 4 \pi R^2 \rho_w v_w
\ee

At times of few months - years the evolution of the spin-down luminosity  $L_{sd}$ of the central \NS\ should be taken into account,
\be
L_{sd} =  \frac{L_0}{(1+t/t_\Omega)^2} 
\ee
where  $L_0$ is the initial spindown power, $t_\Omega $ is the spindown time.   

Neglecting the  swept-up momentum and energy of the preceding wind,  in the  Kompaneets approximation  \citep{1960SPhD....5...46K,1995RvMP...67..661B}  the  non-relativistic source-powered  forward shock  propagates according to 
\be 
\frac{L_{sd} }{4\pi R_{s}^2 c} = \rho_w ( \partial_t R_{s} - v_w)^2
\ee
where  $R_{s}$ is the location of the shock (more precisely, this is the location of the structure  of the reverse shock-contact discontinuity - forward shock in the thin shell approximation; this overestimates the location of the reverse shock). 
 Formally this is a condition at the contact discontinuity, which is located close to the FS. \citep[In passing we note that conclusion of][that absence of overall relativistic expansion  excludes long-lived relativistic outflow is incorrect: overall expansion traces the  post reverse shock flow, not the fast upstream wind, Fig. \protect\ref{veclocity}. Like in PWNe,  the wind is highly relativistic, while the overall expansion is non-relativistic.]{2020MNRAS.491.4735B} 
 
 In the thin shell approximation the reverse shock-contact discontinuity- forward shock system move according to 
\ba  &&
   R_{s}= v_w t + \left(\frac{L_0 v_w}{ c \dot{M}} \right)^{1/2}  t_\Omega \ln (1+t/t_\Omega) \to \left( 1+ \left(\frac{L_0 }{ c v_w \dot{M}} \right)^{1/2} \right) v_w  t  \approx
2 \times 10^{17} \, {\rm cm} \,  l_{0, 43}^{1/2} \dot{M}_{-4} ^{-1/2} v_{w,4}^{1/2}
   \nn && 
   V_s =  \partial_t R_s - v_w =  \left(\frac{L_0 v_w}{ c \dot{M}} \right)^{1/2} \frac{1}{1+t/t_\Omega} \to  \left(\frac{L_0 v_w}{ c \dot{M}} \right)^{1/2} 
   \nn &&
\beta_{s} =   \frac{V_s}{c} = \left(\frac{L_0 v_w}{ c^3 \dot{M}} \right)^{1/2} = 0.2 l_{0, 43}^{1/2} \dot{M}_{-4} ^{-1/2} v_{w,4}^{1/2}
\label{betaFS}
   \ea
   where $l_{0, 43}= L_0/(10^{43}$ erg s$^{-1}$) (this is fixed by the late time X-ray power, see below) and numerical estimates are at $t=328$ days.
   Thus the shock is mildly relativistic (we also assumed $t \leq t_\Omega$, see below). This is consistent with the limit derived by \cite{2020MNRAS.491.4735B}. 

Strong shock condition requires
\be
L_{sd} \geq c v_w \dot{M} = 2 \times 10^{41} \, \dot{M}_{-4} v_{w,4}\, {\rm erg \,s}^{-1}
\ee
This is  satisfied by our fiducial parameters.
 
\subsection{Properties of the central \NS} 
 
In a PWN paradigm \citep[\eg][]{1984ApJ...283..710K,2020ApJ...896..147L},  Fig. \ref{FBOTs-New-RS-X},  the X-ray emission occurs in fast cooling regime: hence all the energy put into particles is radiated. The observed X-ray luminosity $L_X$  is then a fraction $ \epsilon _e$ of the spin-down luminosity $L_{sd} $
\ba &&
L_X = \epsilon _e L_{sd}
\nn &&
L_{sd} \sim B_{NS}^2 R_{NS}^2 c  \left( \frac{ \Omega _{NS} R_{NS} }{c} \right)^4
\label{Lsd} 
\ea
Where $B_{NS}$ is the surface \Bf, $R_{NS}$ is \NS\ radius, and $ \Omega _{NS} $ is the spin of the central \NS.

Scaling the surface \Bf\ with  the quantum field, 
\ba &&
B_{NS}  = b_q B_Q
\nn &&
B_Q = \frac{c^3 m_e^2}{e \hbar }=4\times 10^{13}\, {\rm Gauss},
\ea 
and given the observed X-ray luminosity $L_X$, Eqns (\ref{Lsd}) give 
the required period of a \NS:
\be
P_{NS} = 2 \pi b_q^{1/2}  \epsilon _e^{1/4}  \frac{ B_Q^{1/2 } R_{NS}^{3/2} }{c^{3/4} L_X^{1/4}} = 10b_q^{1/2}   \epsilon _{e,-1}^{1/4} L_{X, 42}^{-1/4}\,  {\rm msec}  
\ee
The corresponding spin-down time
\be
t_\Omega = b_q^{-1/2}   \epsilon _e^{1/2}  \frac{c^{3/2} I_{NS} }{ 2 B_Q L_X^{1/2} R_{NS}^3 }  =  215 \,  \epsilon _{e,-1} ^{1/2}  b_q^{-1} L_{X,42}^{-1/2}\, {\rm days} 
\label{tOmega} 
\ee
A surface field just below the quantum field, $b_q \sim  0.5$ is required to have spindown time longer than the time of Chandra observations at 328 days.

The initial rotational energy of the NS is high, but not extreme:
\be
E_{NS} = L_{sd} t_\Omega= \frac{1}{2 \sqrt{\epsilon_e b_q } }\frac{ c^{3/2} I_{NS} L_X^{1/2} } {B_Q R_{NS}^3} = 1.8 \times 10^{50} b_q^{-1}  L_{X,42}^{1/2} \epsilon_{e,-1} ^{-1/2} \, {\rm erg}
\label{ENS}
\ee
As a check, rotational energy of the \NS\ is somewhat larger than the expected ejecta energy, $E_{NS}  \geq E_{ej}$, Eq. (\ref{RR}). 

Thus, the requirement on the  properties of the central source from X-ray observations are fairly mild: central \NS\ with surface  fields  below the quantum field, spinning at $\sim 10 $ milli-seconds ({\it not} a millisecond magnetar).

\subsection{X-rays: emission from the relativistic   termination shock: power, frequency and variability}

As we discussed in the Introduction, it is  the late X-ray emission that is the most energetically demanding, both in terms of the photon energy and the emitted power. Explaining late X-ray emission is most challenging: even the most energetically powerful, collimated and relativistically beamed  GRBs  typically do not produce X-ray emission after a year.

Following 
\cite{2019MNRAS.487.5618L}, we outline  a PWNe-like picture for the high energy emission,  Fig. \ref{FBOTs-New-RS-X},   that the X-ray emission is generated at the termination shock of a long lasting relativistic wind  \citep[see also][with  applications to GRBs proper]{2011MNRAS.411..422L,2017ApJ...835..206L,2021ApJ...907..109B}.

Unlike the case of emission from the FS, where \Bf\ needs to be amplified, at the termination shock the \Bf\ is supplied by the central source.
At the location of the reverse shock the \Bf\ in the plasma associated with the second, NS-generated, wind is
\ba && 
L_{sd}  = B _{RS}^2 R _{RS}^2 c
\nn &&
B _{RS} = \frac{\sqrt{ L_{sd}  /c} }{R_{RS}} = \frac{ \sqrt{\dot{M} }} { \sqrt{ v_w } t}
\label{11}
\ea
where we used  $R_{RS} \sim \left(   {L_0  v_w }/({ c  \dot{M}} ) \right)^{1/2} t$, Eq.  (\ref{betaFS}),  for the location of the shocks and approximated $L_{sd} \sim L_0$.  At time of 328 days the shocks are located at $\sim 10^{17}$  cm, see (\ref{betaFS}),  with \Bf\ $\sim 0.1$ Gauss.
\citep[This parameterization formally  assumes mild magnetization $\sigma$, but it is in fact applicable to high-$sigma$ flows, as demonstrated by the resolution of the $\sigma$-paradox by ][]{2013MNRAS.431L..48P}.

 The estimate of the properties of the RS (\ref{11}) formally matches the estimate at the  forward shock  (\ref{01})  for $\epsilon_B \sim 1$ 
 \citep[$\epsilon_B$ is a common parameter in the GRB/SNR  theory,][the ratio of  the post-shock amplified magnetic field energy density to the shocked plasma energy density]{2004RvMP...76.1143P}
The main difference  in the  forward shock and RS emission is the typical \Lf\ of the accelerated particles: in the PWN paradigm it is determined by the \Lf\ of the wind, not the energy flux.
If relativistic second wind has \Lf\ $\Gamma_w$, estimating the particles' post-RS random \Lf\  as $\gamma \sim  \Gamma_w$, the peak frequency is 
\be
\epsilon_{\rm X} \approx  \hbar \gamma^2 \om_B  \approx  \hbar  \Gamma_w^2  \left( \frac{ e B _{RS} }{m_e c} \right) =
   \Gamma_w^2  \frac{ e  \hbar  \sqrt{\dot{M}}}{ c m_e \sqrt{v_w} t} =  1 \, {\rm  keV} \, \Gamma_{w,6} ^2  \dot{M}_{-4} ^{1/2} v_{w,4} ^{-1/2}  \left( \frac{t}{328 {\rm days}} \right) ^{-1}
   \label{epsilonX}
\ee
(Like in PWNe, the post-RS flow is non-relativistic in our frame, hence no bulk \Lf\ is involved.)

The required \Lf\ $\Gamma_w \sim 10^6$ is somewhat higher than is usually assumed for pulsars. For example in Crab the termination shock is located at similar distance of $\sim  10^{17}$ cm, while \Bf\ is three orders of magnitude smaller; with the wind's \Lf\ of $\Gamma_w \sim 10^4$ the peak emission falls into near IR \citep[\eg][]{2019MNRAS.489.2403L}. But recall that the thin shell approximation overestimates the location of the reverse shock (hence underestimates the value of \Bf). In addition, high-sigma flows will have reverse shock at smaller radii. All these effects will lead to less strict requirements on the \Lf\ of the wind. In addition, acceleration at the termination shock may be more efficient, \eg\ to additional effects of reconnection \citep[\eg][]{2011ApJ...741...39S}. These points are also important for the variability, see below.

Since the  synchrotron cooling time at the RS is short,
\be
\tau_{\rm cool}  = \frac{ m_e^3 c^5 v_w t^2}{\Gamma_w e^4 \dot{M} } = 4\times 10^3 \, {\rm sec}  \Gamma_{w,6} ^{-1}   \dot{M}_{-4} ^{-1} v_{w,4}   \left( \frac{t}{328 {\rm days}} \right) ^{2}
\label{tau} 
\ee
All the energy of the wind that is injected into accelerating particles is emitted, Eq. (\ref{Lsd}).

Another important observational fact is fast, 
erratic intra-day variability of  the X-ray emission \citep{2019ApJ...871...73H,2019ApJ...872...18M}.
It is hard to reproduce  fast variability within the forward shock scenario since the forward shock emission properties depend on the {\it  integrated} quantities - central engine total energy and total matter swept. But fast variability  can be reproduced within the internal shock paradigm, especially in the highly magnetized winds  \citep{2019MNRAS.487.5618L}. First,  at the termination shock particles emit in the fast cooling regime, (\ref{tau}), hence any variation of the plasma parameters are reflected  in the  emission \citep[as advocated by ][for flares and sudden changes observed in GRB afterglows]{2017ApJ...835..206L}.

Second,   variations of the plasma properties  need not be global, but can be  local. Examples include: (i) variations of Crab wisps \citep{2008ARA&A..46..127H} due to the non-stationarity of the termination shock;  (ii) Crab flare-like reconnection processes in the shocked pulsar wind \citep{2012MNRAS.426.1374C,2018JPlPh..84b6301L}; (iii) anisotropic local  emission \citep[``jet-in-jet'' model of ][the latter works especially well in the exhaust jets of highly magnetized reconnection regions]{Lyutikov:2006a,2010MNRAS.402.1649G}.

\subsection{Radio-mm    emission  from  the forward shock in the wind}

Self-consistent models 
of radio emission of astrophysical sources is a notoriously difficult problem, partly because radio emission is energetically subdominant \citep[\eg the problem of radio emission of Crab PWN,][]{1984ApJ...283..710K}. 
Microphysics of electron  injection at shock is most complicated
\citep{2011ApJ...733...63R}. In the case of highly magnetized winds, a combination of Fermi and reconnection processes are likely at play \citep{2014ApJ...783L..21S,2019MNRAS.489.2403L}. 

A simple approach is to use 
the classic GRB prescription \citep{Sari95,Sari97}, see also \cite{2017hsn..book..875C}.
The peak  \Lf\ of  particles accelerated  at the   forward shock propagating into  unmagnetized medium  with velocity $\beta_{FS}$  is expected  to be 
\be 
\gamma_{FS} \sim  \frac{m_p}{m_e}  \beta_{FS}^2 =  \frac{m_p}{m_e}  \frac {  L_{sd}  v_w}{\dot{M} c^3}
\label{02} 
\ee
with $\beta_{FS}$ given by (\ref{betaFS}).


In the preceding wind  the \Bf\ is weak, and needs to be amplified to produce synchrotron emission. 
The  post- forward shock energy density
\be
u_{FS} \sim \rho_w \beta_{FS}^2  c^2 = \frac{L_{sd} } {4  \pi R^2 c} = \frac{\dot{M}}{ 4\pi v_w t^2}
\ee 
gives the \Bf\
\be
\frac{B_{FS} ^2}{8\pi} =\epsilon_B  u_{FS} \to B_{FS}  = \sqrt{2 \epsilon_B} \frac{\sqrt{L_s}}{ R \sqrt{c} } = \sqrt{2 \epsilon_B} \frac{\dot{M} }{ \sqrt{v_w} t}
\label{01} 
\ee

The \Bf\  (\ref{01}) and post- forward shock \Lf\   (\ref{02})  give the 
emission frequency
\be
\frac{\om}{2\pi}  \sim \gamma_{FS} ^2 \frac{ e B_{FS}}{ 2 \pi m_e c} =
\frac{\sqrt{ \epsilon_B} }{ \sqrt{2} \pi  }
 \frac{ e m_p^2}{c^7 m_e^3} \frac{L_s^ 2 v_w ^{3/2} } {\dot{M} ^{3/2} t}  
=1.4  \epsilon_{B,-1}^{1/2}  l_{0,43} ^2 v_{w,4}^{3/2} \dot{M}_{-4}^{-3/2}  \left( \frac{t}{ 261{ \rm days}}\right) ^{-1} \, {\rm GHz} 
\label{nuradio} 
\ee
at 261 days.
This matches the value of the peak emission at  261 days, and decrease with time \citep[Fig.  7 of][]{2021arXiv211200751Y}. The peak emission frequency  (\ref{nuradio}) is a sensitive function of the central source's luminosity, pre-explosion wind parameters and time of observations.

The above estimates also exclude forward shock as the origin of the X-ray emission. Though the injection problem at shocks is notoriously difficult, the estimates are off by  $\sim  10$   orders of magnitude in emitted frequency,    $\sim  5$   orders of magnitude in particle energy.
 Recall, that the late X-ray emission is the dominant energy channel in AT2020mrf. To bring emission energy to the  keV  range, the \Lf\ of the X-ray emitting particles should exceed the peak \Lf\ (\ref{02})  by  factor $\sim 10^5$. In that case the total energetics is higher by a similar factor (for distribution  function $f \propto \gamma^{-2}$).

\section{Contribution from radioactivity}
\label{Contributionfromradioactivity}
The  envelope ejected during AIC  will contain some $^{56}$Ni; up to  $\sim$ few times $10^{-2}$ of  $^{56}$Ni  are expected \citep{1999ApJ...516..892F}.
 This will not affect the FBOT light   curve (half- life time of $^{56}$Ni  is 6.1 days), but  this does affect the longer light curves. For example, a decay of $10^{-2} M_\odot$ of     $^{56}$Ni  ejected would produce (at $3.6$ MeV emitted as radiation per decay), total \EM\ energy $1.3 \times 10^{48}$ erg, comparable to the overall energetics (if reprocessed to optical).
 
 An important issue here is that for small ejecta masses the radiative energy is more efficiently converted into radiation, as it is not degraded by adiabatic  losses in the typical  months-long optically thick stage of expansion. The ejecta becomes fully transparent at  time (\ref{tau23}), just somewhat longer than the  $^{56}$Ni decay half life. 
Hence, unlike the conventional SNIa where most of the   $^{56}$Ni energy is ``wasted''  into the kinetic energy of
expansion, a larger fraction is radiated now.

\section{Transients following WDs mergers}
\label{Transits}

Mergers of WDs is one of the most frequent catastrophic events \citep[][]{2012ApJ...748...35S,2016MNRAS.463.3461S}. Supernova of Ia type is the most frequently discussed channel  of WDs mergers \citep[][]{2014ARA&A..52..107M,2019NewAR..8701535S}.  The rates of SNIa are much smaller than of the WD  mergers \citep{2017MNRAS.464.1607Y}.Mergers of CO-CO WDs with a combined mass above Chandrasekhar, the double-degenerate  progenitors model  for supernova type Ia  \citep{1984ApJ...277..355W, Ibe84}, have a merger rate of $1.7-2.2\times 10^{-3}$yr$^{-1}$ per Galaxy. This is about an order of magnitude above that of CO-ONeMg WDs rates, estimated
by \cite{2019MNRAS.487.5618L}  to be   few $\times 10^{-4}$yr$^{-1}$.  The estimated CO-ONeMg WDs merger rate is consistent with the lower limit of the FBOT rate.  

We suggest that FBOTs is one of the results of the   super-Chandrasekhar double WD merger. FBOTs correspond to 
 a narrow parameter range in the pre-merger WDs masses and compositions, and possibly spins (and correspondingly narrow parameter range in the main sequence masses and separations).

There is a number of possibly related phenomena. First, 
the picture  of \cite{2019MNRAS.487.5618L}  (that the merger of a heavy ONeMg WD with a CO WD   creates a super-Chandrasekhar mass,  shell  C and O burning star with  luminosity  $L\sim 10^4 L_\odot$ with strong winds) 
was well matched by  observations  of \cite{2019Natur.569..684G,2020A&A...644L...8O} of  IRAS 00500+6713:   a hot, $\sim 200,000$K, luminous   $\sim10^{4.5} L_\odot$ star within a mid-infrared nebula. Both the star and the nebula appear to be free of hydrogen and helium. The wind velocity is exceptionally high, $\sim 16,000$ km s$^{-1}$.
The central star and the nebula  are thus the system created by a WD merger and may produce FBOT in the future.

A highly magnetized and rapidly rotating white dwarf as small as the Moon \citep{2021Natur.595...39C}  is the remnant of the merger. It was on the verge of  electron-capture collapse, but run out of the material in the envelope and remained intact.

Next we mention a few works that are contributing to various arguments given above. 
 \cite{2012MNRAS.427.2057H}   proposed a model  of super-Chandrasekhar explosion dominated  by the interaction of the SN ejecta with a hydrogen- and helium-poor circumburst medium. In our model it's the interaction with the pre-explosion H-poor progenitors wind.
Another possibly relevant work is by  \cite{2021ApJ...923L..24S}  who discuss a population of  radio-luminous supernovae.  \cite{2009A&A...500.1193L} found that during the merger itself very little mass, $\sim 10^{-3} M_\odot$, is ejected hydrodynamically. If AIC occurs $\geq 100$ yrs after the merger, that primary ejected shell is reached by the AIC ejecta/shock on time scale of a year.   \cite{2014MNRAS.438.1005D} argued that the pulsar J0737-3039B was a product of an AIC of ONeMg WD with small shell mass; in our model this would require that the initial system was a triple.

Finally, there are  possibly related Type Iax class of supernovae  \citep[subluminous Ia-s,][]{2013ApJ...767...57F}  and  Type Ibn  \citep[][]{2017ApJ...836..158H}.  Type Iax may  be the results of AICs with larger ejecta mass $M_{ej} \geq 0.1$;  Type Ibn may be the result of merger with a He WD. \citep[Somewhat similarly][argued that the ejecta mass is the dominant  factor affecting the light curves ]{2015MNRAS.452.3869N}
 We live investigation of these possibilities to a future research.

\section{Conclusion and prediction}
\label{Conclusion}

The present model of broad band emission from FBOTs  is highly constrained  - it reproduces in a self-consistent manner the  optical FBOTs, late radio-mm  emission and the dominant  long term X-ray emission.  One of the key advantages of the model is the low requirement on energetics, Eq. (\ref{ENS}), only few $\times 10^{50}$ ergs,   just a bit higher than the observed X-ray energy. This is a consequence of small ejecta mass. This low energy budget compares favorably  with GRB-like energy budgets of $\geq 10^{52}$ ergs in  models that advocate large ejecta mass 
\citep[\eg][]{2022MNRAS.513.3810G,2022arXiv220112534C}.  Our inferred rotational period of few milliseconds is consistent with possible detection of periodicity by \cite{ 2022NatAs...6..249P}.

There are several principal points we make. First,  the present  model naturally explains small ejecta mass. This is due to the competition of envelope mass lost to the wind, and ashes added to the core. FBOTs results from AIC when the shell mass is $\leq 0.1 M_\odot$. If AIC occurs with larger shell mass then the shell remains optically thick longer and most of the energy dissipated at the \NS-driven shock is not radiated, but is spent on $pdV$ work. FBOTs thus correspond to a narrow range of initial parameters of the binary system -  this also explains their rarity.  FBOTs' emission is generated as the  radiation-dominated  forward shock propagates though ejecta.

The second major point is that the overall energetics is dominated in the case of AT2020mrf by  long-term X-ray emission. Both the production of X-rays at times $\sim $ one year, and the total energetics are the most demanding. Even the most energetic  GRBs  do not produce X-ray that late.  We outlined a  PWN-type model of high energy emission, where X-rays are  generated at the relativistic termination shock (not the forward shock). The demands on the central \NS\ are not extreme - a combination of high \Bf\ and fast spins (\eg 10 milliseconds)  is needed,  but  not a millisecond magnetar with quantum field. Both the early and  late X-rays are produced at the NS-driven termination shock: first inside the ejecta, and later in the preceding wind. Since X-ray emission is in the fast cooling regime (high \Bfs\ in the wind) a large fraction of the wind energy is radiated. 

Third, the duration of the FBOTs is determined by the diffusion of photons produced deep in the ejecta, in the regions of high optical depth, Eqns. (\ref{tFBOT}-\ref{tFBOT1}). This is somewhat new regime, as typically  supernova light curves were determined mostly by distributed source of photons (bremsstrahlung, recombination and radioactive decay). In our case a powerful, highly non-equilibrium  source of photons is located  deep in the ejecta. 

In the case of centrally produced photons, within a high optical depth ejecta, nearly all the photons escape at the same time (assuming spherical symmetry of the ejecta -  a far from certain assumption). Thus, the peak flux is an integrated property: all the photons produced by the NS shock before $t_{FBOT}$ escape within a narrow time near  $t_{FBOT}$. 

Another constraint comes from the requirement that the  NS-generated shock should break through ejecta to produce the SRG/eROSITA X-ray bump. For this the energy deposited by the NS should be larger than the energy of the ejecta. Sufficiently light ejecta, and sufficiently powerful NS  are needed. SRG /eROSITA X-ray bump  corresponds to the   break-out of the shock from the ejecta into the preceding wind; this was predicted by  \cite{2019MNRAS.487.5618L}.

The present  model explains
\begin{itemize} 
\item  short rise times of FBOTs (light ejecta)
\item blue color (radiation-dominated shock)
\item decreasing photospheric radius \citep{{2019MNRAS.484.1031P}} 
\item change of properties at $\sim$  1 month (shock exiting the ejecta into the wind)
\item   predicted the X-ray bump seen by eROSITA in AT2020mrf (due the shock breakout)
\item long late X-ray emission (from the termination shock)
\item long radio-mm (from the forward shock) 
\item presence of dense circumburst medium  \citep[inferred by][]{2022ApJ...926..112B,2021ApJ...912L...9N} - this is the post-merger/pre-collapse wind
\item Host galaxies: as discussed by \cite{2019MNRAS.487.5618L}, their Fig. 6, the merger rates   of the CO-ONeMg  WDs peaks at short delay times of about $\sim$ 50-100Myr, with a long tail to long delay times.  Thus the typical delay time of the CO-ONeMg mergers is closer to that of core-collapse supernovae.  \cite{2019MNRAS.487.5618L} expected  the host galaxies of CO-ONeMg mergers to be more similar to those of   core-collapse supernovae instead of Type Ia supernovae. This explains detection of FBOTs preferentially in star-forming galaxies \citep{2018ApJ...865L...3P,2020MNRAS.495..992L}. We expect further detections in the early-type galaxies though.
\end{itemize}

The model has a number of predictions: 
\begin{itemize} 
\item the system should be hydrogen poor, but not necessarily  hydrogen free. If the secondary WD was of the DA type, the most common,   the  ejecta and the wind will likely have   $\sim 10^{-4} M_\odot$ of hydrogen. \cite[][noted that H-poor spectra of  SN-Ibn  are similar to that of AT 2018cow at later times]{2019MNRAS.488.3772F,2021arXiv211015370P}. An advanced radiation transfer modeling is required.
\item one expects evidence (\eg spectral) of shock interaction with  fast and dense pre-explosion wind 
\item anisotropy is expected: both the ejection of the envelope during AIC, as well as  \NS-driven winds are expected to be anisotropic. 
\item pre-FBOT  archival data should show a  bright persistent   hydrogen-poor  source, possibly surrounded by a nebula \cite[similar to  IRAS 00500+6713,][]{2019Natur.569..684G,2020A&A...644L...8O} 
\item in the case when AIC does not occur,   a hydrogen poor envelope around the central (massive)   WD  is expected. It's observability depends on how long ago the last part of the envelope was lost.  In the case of  massive WD observed by  \cite{2021Natur.595...39C},  the age is few million years, so the envelope is dissipated by now.
\end{itemize}

\section*{Acknowledgments}

I would like to thank Igor Andreoni, Ilaria Caiazzo,  Paul Duffel, Ori Fox,  Dimitrios Giannios,   Daniel Kasen, Danny Milisavljevic, Lida Oskinova, John Raymonds,  Noam Soker, Bhagy Subrayan,  Silvia Toonen,  Alexander Tutukov, Beatriz Villarroel for discussions.  Comments by late Vasilii Gvaramadze are acknowledged.   This work had been supported by 
NASA grants 80NSSC17K0757 and 80NSSC20K0910,   NSF grants 1903332 and  1908590.


\section{Data availability}
The data underlying this article will be shared on reasonable request to the corresponding author.

\bibliographystyle{apj}

  \bibliography{/Users/maxim/Home/Research/BibTex}

\appendix
\end{document}